\documentclass[twocolumn,prb,amsmath,amssymb,amsfonts,superscriptaddress,floatfix,aps,showpacs]{revtex4-1}

\usepackage{graphicx}
\usepackage{dcolumn}
\usepackage{bm}
\usepackage{rotating} 
\usepackage{color}
\usepackage{subfigure}
\usepackage{natbib}
\begin{document}
\newcommand{\red}{\color{red}}

\title{Treatment of 4f states of the rare-earths: the case study of TbN}
\date{\today}
\author{L. Peters}
\affiliation{
Institute for Molecules and Materials, Radboud University Nijmegen, NL-6525 AJ Nijmegen, The Netherlands
 }
\author{I. {Di Marco}}
\affiliation{
Department of Physics and Astronomy, Uppsala University,
Box 516, SE-75120, Uppsala, Sweden
 }
\author{P. Thunstr\"om}
\affiliation{
Institut fur Festkorperphysik, Vienna Univeristy of Technology, 1040 Wien, Austria
 }
\author{M.I. Katsnelson}
\affiliation{
Institute for Molecules and Materials, Radboud University Nijmegen, NL-6525 AJ Nijmegen, The Netherlands
 }
\author{A. Kirilyuk}
\affiliation{
Institute for Molecules and Materials, Radboud University Nijmegen, NL-6525 AJ Nijmegen, The Netherlands      
 }
\author{O. Eriksson}
\email{olle.eriksson@physics.uu.se}
\affiliation{
Department of Physics and Astronomy, Uppsala University,
Box 516, SE-75120, Uppsala, Sweden
 }

\begin{abstract}
The lattice constant, bulk modulus and shear constant of TbN are calculated by means of density functional theory (DFT) in the local density approximation (LDA) and generalized gradient approximation (GGA), with 4f-states treated as valence electrons or core electrons. In addition, local Coulomb repulsions U are treated both statically as in the LDA+U approach and dynamically as in the dynamical mean-field theory (DMFT) in the Hubbard-I approximation. It is shown that all methods, except DFT-LDA with 4f electrons treated as either valence states, produce lattice constants and bulk moduli in good agreement with experiment. In the LDA+U approach multiple minima are found, and we focus on the competition between a state with cubic symmetry and a state obtained from atomic Hund's rules. We find the state with cubic symmetry to be 0.59 eV lower in energy than the Hund's rules state, while the opposite was obtained in previous literature. The shear constant is shown to be rather sensitive to the theoretical method used, and the Hund's rules state obtained in LDA+U is found to be unstable towards tetragonal shear. As to the magnetism, we find that the calculation based on the Hubbard-I approximation reproduces observations with the best accuracy. Finally, the spectral properties of TbN are discussed, together with the general applicability of the different methods in describing rare-earth elements and compounds.
\end{abstract}
\maketitle
\noindent

\section{Introduction}
In the past decades, it has been shown that calculations based on density functional theory (DFT)~\cite{hohenberg64pr136:B864,kohn65pr140:a1133} reproduce measured materials properties, e.g. the elastic constants, equation of state, catalytic activity, conductivity, lattice dynamics, surface tension, work function and the spin- and orbital moments, with good accuracy for most elements and various compounds~\cite{eriksson_encyclopedia_06}. This conclusion holds for those systems with weak electron-electron correlations, where the exchange correlation functional can be parametrized using information from the uniform electron gas as in the local density approximation (LDA) or generalized gradient approximation (GGA). This, however, is far from the situation of the 4f shell of the rare-earth elements, where the direct electron-electron repulsion is significant and can not without further effort be incorporated in {\it ab-initio} theory, where no input is expected from experimental data.

Based on the wealth of experimental information available for the rare-earths~\cite{*[{See articles in: }] [{}] {hpcre_78}}, it is by now established that the 4f shell has localized electron states, where band-dispersion effects are negligible. The electron-electron repulsion within the 4f shell is found to be minimized by the formation of a Russell-Saunders coupled ground state, and with the exception of the $\alpha$-phase of Ce, Eu, and Yb, all rare-earth elements form a trivalent configuration in the solid. Eu and Yb are divalent, since this configuration provides a half-filled or filled 4f shell~\cite{johansson79prb20:1315}. This understanding of the 4f shell of the rare-earths is the basis of the so called Standard Model of this class of elements~\cite{jensen_book}. The Standard Model explains all the essential properties of the rare-earth elements, like the crystal structure~\cite{duthie77prl38:564,skriver85prb31:1909}, equilibrium volume~\cite{skriver85prb31:1909,delin98prb58:4345}, bulk modulus~\cite{skriver85prb31:1909,delin98prb58:4345}, valence stability~\cite{delin97prl79:4637}, crystal field splittings and the magnetic phase diagram~\cite{nordstrom00epl49:775}. The Standard Model applies as well to compounds involving rare-earth elements, albeit in some cases a mixed valent behaviour is observed~\cite{*[{See articles in: }] [{}] {vfs_book,vi_book}}, where the electronic configuration of the 4f shell fluctuates between two integer occupancies, i.e. f$^n$ and f$^{n+1}$.

Any theory on the electronic structure of the rare-earth elements should reproduce the behaviour observed in the Standard Model. In the past this was achieved by treating the 4f electrons as being part of the core-states, so that measured densities, structural stability and bulk modulus were reproduced with good accuracy~\cite{duthie77prl38:564,skriver85prb31:1909,delin98prb58:4345}. The inter-atomic exchange interaction, which is in this case given by the RKKY mechanism, was also reproduced by a theory that treats the 4f electrons as part of a chemically inert core~\cite{nordstrom00epl49:775}.
Lately, parametrized Hartree-Fock theory in the form of the LDA+U approximation~\cite{anisimov91prb44:943} has become popular for treating the electron-electron repulsion of the 4f shell~\cite{anisimov97jpcm9:767,antonov01prb63:205112,larson07prb75:045114}. Although the chemical inertness of the 4f shell can be achieved in this way, by pushing occupied states to low energies, and unoccupied states well above the Fermi level, it is unclear how well the calculated electronic structure agrees with measured valence band spectra. It is also not clear whether the LDA+U approximation can reproduce more delicate materials properties of rare earths, like elastic constants, magnetic moments, or valence stability. 

Dynamical mean field theory (DMFT)~\cite{*[{For a review see: }] [{}] kotliar06rmp78:865} in the form of the Hubbard-I approximation (HIA)~\cite{lichtenstein98prb57:6884} has recently shown promising results in describing the spectral properties of several rare-earth systems~\cite{svane06ssc140:364,lebegue05prb72:245102,pourovskii07prb76:235101,thunstrom09prb79:165104,pourovskii09prl102:096401,litsarev12prb86:115116}. The treatment of the 4f shell in this way, holds great promise since it naturally describes many of the experimentally known facts of the rare-earths, in particular the Russell-Saunders ground state and the formation of atomic multiplets.

In this work we apply the theories discussed so far for the rare-earths, to the terbium nitride compound. These theories will be compared for the calculation of the lattice constant, bulk modulus, shear constant, magnetic moments and one-particle excitation spectrum. Terbium nitride was chosen as it is a particularly significant example of the interplay between atomic-like effects and one-electron crystal field splittings, which provides a complication for effective one-electron theories. Moreover, TbN, and all other rare-earth nitrides are very relevant for the scientific community, due to the easily tunable magnetic properties, which often coexist with a semiconducting character, making them interesting candidates for spintronics~\cite{natali13pms58:1316}.

\section{Details of Calculations}
All the calculations reported in the present paper were carried out using a full potential linear muffin-tin orbital (FP-LMTO) method~\cite{RSPt_book}. We used LDA and GGA parametrizations of the exchange-correlation functional as formulated by Perdew and Wang~\cite{perdew92prb45:13244} and by Perdew, Burke, and Ernzerhof~\cite{perdew96prl77:3865}. The Brillouin zone was sampled through a conventional Monkhorst-Pack mesh of 30 x 30 x 30 {\bf{k}}-points, leading to a total of 904 vectors in the irreducible wedge. The basic geometrical and basis setup was the same for all calculations, with the exception of the 4f-states, described below. For the definition of the muffin-tin sphere of nitrogen we used a radius of 2.056 a.u., and for terbium one of 2.18 a.u. in case of LDA and 2.41 a.u. for GGA. This smaller radius for LDA was necessary due to the overbinding tendency of LDA with the 4f-electrons in the valence (see Table~\ref{structure1}). The main valence basis functions were chosen as 6s, 6p and 5d states, while 5s and 5p electrons were treated as pseudocore in a second energy set~\cite{RSPt_book}. The 4f-states were treated as valence states for some simulations and as core states for some other simulations. In the latter case 5f-states were instead added to the valence electrons, in order to have basis functions with f angular character. Three kinetic energy tails were used for 6s and 6p states, corresponding to the default~\cite{RSPt_book} values 0.3, -2.3 and -0.6 Ry. Only the first two tails were used for all the other basis functions.

Apart from pure DFT in LDA or GGA, we also performed simulations in combination with DMFT~\cite{kotliar06rmp78:865}. Details on the implementation used in this work are given elsewhere~\cite{grechnev07prb76:035107,dimarco09prb79:115111,thunstrom09prb79:165104,granas12cms55:295,thunstrom12prl109:186401}, and we refer the reader to those studies for a complete description of our methods. In the present paper the effective impurity problem arising in the DMFT cycle was solved in the HIA~\cite{thunstrom09prb79:165104}. Conforming to existing notation, we will address this method with the acronym LDA+DMFT[HIA]. Moreover we have performed other calculations, where the effective impurity model was solved in the Hartree-Fock approximation, which corresponds to the LDA/GGA+U approach~\cite{anisimov91prb44:943,anisimov97jpcm9:767} in the most general fully rotationally invariant form~\cite{kotliar06rmp78:865}. In the LDA/GGA+U and LDA+DMFT[HIA] simulations we used slightly different local orbitals to which we applied the Hubbard U correction, respectively ORT and MT orbitals. 
These orbitals are constructed from LMTOs, that have a representation involving structure constants, spherical harmonics, and a numerical radial representation inside the muffin-tin spheres. These functions are matched continuously and differentiably at the border of the muffin-tin spheres to Hankel or Neumann functions in the interstitial. The ORT basis originates from these native LMTOs after a L\"owdin orthonormalization. The MT orbitals, instead, are atomic-like orbitals where the radial part comes from the solution of the radial Schr\"odinger equation inside the muffin-tin sphere at an energy corresponding to the 'center of gravity' of the relevant energy band. For a more detailed description about the correlated orbital bases we refer to Ref.~\onlinecite{grechnev07prb76:035107}. 
There, it is also shown that they generally lead to very similar results. Finally the double counting correction~\cite{kotliar06rmp78:865}  was set up in the fully localized limit (FLL)~\cite{anisimov97jpcm9:767} for the LDA/GGA+U simulations, while in LDA+DMFT[HIA] was fixed to adjust~\cite{thunstrom09prb79:165104,granas12cms55:295} the position of the first multiplet peak below the Fermi level at -0.15 Ry, which is the measured value for trivalent elemental Tb. This value was kept unchanged for different strains and lattice constants, analogously to what done in Ref.~\onlinecite{litsarev12prb86:115116}. Concerning the Coulomb interaction parameters, a \textit{$U$} of 9.46 eV and a \textit{$J$} of 1.246 eV were used, in agreement with the work of P. Larson et al.~\cite{larson07prb75:045114}. This choice corresponds~\cite{anisimov97jpcm9:767} to the Slater integrals \textit{$F_{0}=9.46$} eV, \textit{$F_{2}=14.97$} eV, \textit{$F_{4}=10.00$} eV and \textit{$F_{6}=7.40$} eV.


In order to obtain the lattice constant and bulk modulus we calculated the total energy for different atomic volumes. These data were fitted through the Murnaghan equation of state~\cite{murnaghan37}, which gave us the equilibrium volume $V_0$ and bulk modulus. For a cubic lattice and small strains, it can be shown that the shear constant, $C^{\prime}$, can be obtained from the expression~\cite{fast_thesis}
\begin{equation}
\frac{\Delta E}{V_{0}}=6C^{\prime}\delta ^2.
\label{STRAIN}
\end{equation}
Here \textit{$\Delta E$} is the total energy difference with respect to equilibrium volume caused by the strain \textit{$\delta$}. This corresponds to a volume conserving strain matrix
\begin{equation}
\begin{pmatrix}
1+\delta & 0& 0\\
0& 1+\delta & 0\\
0& 0& \frac{1}{(1+\delta)^{2}}
\end{pmatrix}
\end{equation}
which acts on the unit cell vectors. The muffin-tin radii were kept fixed for all the calculations for different strains and atomic volumes to minimize numerical errors in the energetics of the core states. 

\section{Results}

\subsection{Lattice properties}
Just like the other rare-earth nitrides, TbN crystallizes in the rocksalt structure~\cite{natali13pms58:1316,larson07prb75:045114}. Equilibrium lattice constants \textit{$a$} and bulk moduli \textit{$B$} obtained with the aforementioned computational methods are reported in Table~\ref{structure1}, together with the experimental values~\cite{hulliger_hpcre_79,wachter98ssc105:675}. All the results presented here were obtained without spin-orbit coupling. This approximation is motivated by the fact that spin-orbit coupling effects are small for the delocalized spd-states and thus should not influence much the bonding properties~\cite{skriver85prb31:1909}.  In the next two subsections we will discuss the effects of spin-orbit coupling more in detail.

\begingroup
\squeezetable
\begin{table}[b]
\begin{tabular}{|l|c|c|c|}
\hline
& & & \\[-2.2ex]
\textbf{Method} & \textbf{\textit{$a$} (\AA)} & \textbf{\textit{$B$} (GPa)} & \textbf{\textit{$C^{\prime}$} (GPa)}\\
\hline
& & & \\[-2.2ex]
LDA \texttt{VALENCE} & 4.77 & 186 & 166 \\
\hline
& & & \\[-2.2ex]
GGA \texttt{VALENCE} & 4.91 & 140 & 115 \\
\hline
& & & \\[-2.2ex]
LDA \texttt{CORE} & 4.90 & 177 & 160 \\
\hline
& & & \\[-2.2ex]
GGA \texttt{CORE} & 4.99 & 162 & 146 \\
\hline
& & & \\[-2.2ex]
LDA+U \texttt{CUBIC} & 4.87 & 179 & 147 \\
\hline
& & & \\[-2.2ex]
LDA+U \texttt{HUND} & 4.87 & 182 & $< 0$ \\
\hline
& & & \\[-2.2ex]
GGA+U \texttt{CUBIC} & 4.97 & 152 & 114 \\
\hline
& & & \\[-2.2ex]
LDA+DMFT[HIA] & 4.89 & 160 & 145 \\
\hline
& & & \\[-2.2ex]
Experiment & 4.92 & 150 & -- \\
\hline
& & & \\[-2.2ex]
Theory from Ref. ~\onlinecite{rukmangad0947:114}& -- & -- & 115 \\
\hline
& & & \\[-2.2ex]
Theory from Ref. ~\onlinecite{ciftci12sss14:401}& -- & -- & 131 \\
\hline
\end{tabular}
\caption{Calculated and experimental values for equilibrium lattice constant, bulk modulus, and shear constant of TbN-bulk. The theoretical values are obtained by means of LDA and GGA for 4f-electrons treated as core states (\texttt{CORE}) and as valence states (\texttt{VALENCE}), whereas for LDA+U and GGA+U solutions with cubic symmetry (\texttt{CUBIC}) and maximal orbital moment (\texttt{HUND}) are reported. LDA+DMFT[HIA] refers to a LDA+DMFT calculation where the effective impurity problem is solved within the Hubbard I approximation. The calculated values are compared with experimental values for the equilibrium lattice constant and bulk modulus~\cite{hulliger_hpcre_79,wachter98ssc105:675}, while two previous computational studies~\cite{rukmangad0947:114,ciftci12sss14:401} are used as reference for the shear constant.}
\label{structure1}
\end{table}
\endgroup
The first column of Table~\ref{structure1} specifies the method used for the calculation, as described in the previous section. The label \texttt{VALENCE} refers to the treatment of the 4f-electrons as valence electrons, while the label \texttt{CORE} indicates that the 4f-electrons are treated as non-hybridizing core states. In case of LDA+U and GGA+U the 4f-electrons are treated as valence states, so none of the previous labels is needed. However, when this method is applied to f-electron systems, a plethora of local minima can be obtained, corresponding to different local density matrix at convergence. Here we consider two significant electronic configurations, labelled as \texttt{HUND} and \texttt{CUBIC}. The former corresponds to a Russell-Saunders coupling of the 4f states, which is consistent with the Standard Model of the rare-earths, while the latter corresponds to the solution where the 4f-configuration respects the cubic symmetry of the lattice. These two solutions are usually found by converging from different starting density matrices. In our calculations, instead, we applied different initial potentials whose symmetries were broken with respect to certain multipole moments~\cite{bultmark09prb80:035121}. At convergence these two approaches are supposed to be equivalent. For GGA+U we report only results for the \texttt{CUBIC} state, since it was not possible to obtain the solution corresponds to the \texttt{HUND} state.

From Table~\ref{structure1} it is clear that all methods except LDA  \texttt{VALENCE} reproduce the lattice constant very well. The bulk modulus appears to be more sensitive to the method used. However, all methods except LDA \texttt{VALENCE} and LDA+U \texttt{HUND} give a value within 20 \% of the experimental value. For the shear constant $C^{\prime}$ there are unfortunately no experimental data available and therefore we compared our calculations with other theoretical analyses~\cite{rukmangad0947:114,ciftci12sss14:401}. The study from Ref.~\onlinecite{rukmangad0947:114} is based on a two-body interionic potential theory with modified ionic charge to include the Coulomb screening effect. The study from Ref.~\onlinecite{ciftci12sss14:401}, instead, is based on DFT through a projector-augmented-wave (PAW) method in GGA. 

All calculations except one lead to a positive shear constant, which indicates that the cubic structure is stable under the considered deformation.  The lack of a positive shear constant for the LDA+U \texttt{HUND} calculation proves that this calculation has an inner instability towards a tetragonal strain. We explored different shears to find the crystal geometry corresponding to the minimal energy in the  LDA+U \texttt{HUND} calculation. We found that a volume conserving strain along the z-direction resulted in the ground state when the c/a ratio was about 0.985.  In Ref.~\onlinecite{larson07prb75:045114} it was argued that the cubic symmetry breaking of the 4f charge density would not have major effects on the measured x-ray diffraction spectra, due to the small contribution to the total charge density. However, our results show that this symmetry breaking produces a sizeable tetragonal distortion of the lattice, which is in contradiction with the experimentally observed cubic crystal structure.

\begin{figure}[b]
\includegraphics[trim=29 390 15 7,clip,width=8cm]{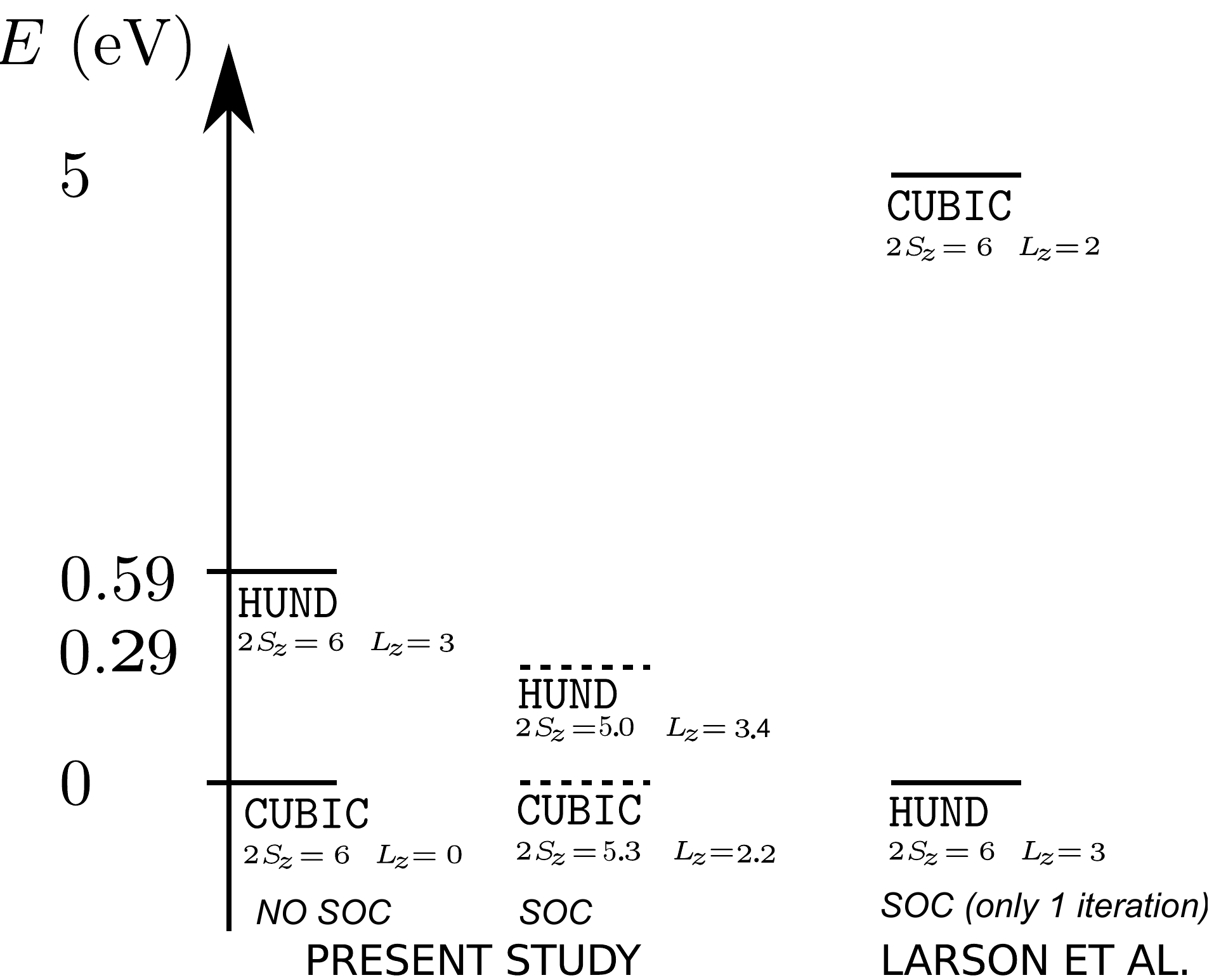}
\caption{Schematic representation of LDA+U total energies, f-projected orbital moment ($L_z$) and spin moment ($2S_z$) of \texttt{CUBIC} and \texttt{HUND} states as calculated in the present paper and by P. Larson et al.~\cite{larson07prb75:045114}. In the 'present study' part the full and dashed lines correspond to respectively a calculation without (\textit{NO SOC}) and with (\textit{SOC}) spin-orbit coupling. In the P. Larson et al. part the total energy difference corresponds to a scalar relativistic calculation. The moments are obtained by doing one iteration with spin-orbit coupling on top of this fully converged scalar relativistic calculation.}\label{Scheme}
\end{figure}

A more detailed comparison of the \texttt{CUBIC} and \texttt{HUND} states in our study and the corresponding states reported by P. Larson et al.~\cite{larson07prb75:045114} is given in Figure~\ref{Scheme}. In this figure the full and dashed lines of the 'present study' part correspond to respectively a calculation without (\textit{NO SOC}) and with (\textit{SOC}) spin-orbit coupling. For the part of this figure corresponding to P. Larson et al. it is important to note that the total energy difference comes from a scalar relativistic calculation without spin-orbit coupling. However, in their study the f-projected orbital and spin moments, respectively $L_z$ and $2S_z$ where $z$ is the magnetization direction, are obtained by turning on spin-orbit coupling for one iteration after converging this scalar relativistic calculation. Thus, for comparing the total energy difference between the \texttt{CUBIC} and \texttt{HUND} states of P. Larson et al. and our study, we should use the results obtained without spin orbit coupling. We find that the LDA+U \texttt{CUBIC} state is more favourable in energy than the \texttt{HUND} state of 0.59 eV. P. Larson et al. find instead the opposite result~\cite{larson07prb75:045114}, and with a much larger energy difference, i.e. about 5 eV. The \texttt{HUND} states in both corresponding studies have the same 4f spin moment (6 $\mu_B$) and orbital moment (3 $\mu_B$). The total moment in this case becomes 9 $\mu_B$, which corresponds well to the total magnetic moment obtained from Russell-Saunders coupling and to what in general is expected for a trivalent Tb atom in elemental form, or in any compound. For the \texttt{CUBIC} states, when comparing our results with those by P. Larson et al., only the 4f spin moment is in good agreement, and has a value of about 3 $\mu_B$. The orbital moments, instead, are different, as reported in Figure~\ref{Scheme}. This is due to the scheme used in Ref.~\onlinecite{larson07prb75:045114} to extract the magnetic moments as explained above. This is why the \texttt{CUBIC} state of P. Larson et al. does not have pure cubic symmetry and has a non-zero orbital moment. In the next subsection (Spin-orbit coupling and magnetic properties) we discuss the orbital and spin magnetic moments coming from a fully self consistent treatment of the spin-orbit coupling. Finally, we would like to emphasize that for the total energy difference between the \texttt{CUBIC} and \texttt{HUND} states, the same configurations are used as in the work of P. Larson et al. For the \texttt{CUBIC} state this means that the minority spin electron occupies the \textit{$a_{2u}$} state and for the \texttt{HUND} state the state with $L_z=3$ quantum number is occupied.    

To further analyse the disagreement in the ground state, we performed additional LDA+U calculations with the full-potential linear augmented plane-wave method (FLAPW) FLEUR~\cite{fleur_website}. Here we found that the \texttt{CUBIC} state is 0.58 eV lower in energy than the \texttt{HUND} state, in accordance to the FP-LMTO results. Finally we should mention that we also explored the effects of the inclusion of an additional term $U_\text{d}$ for the local Coulomb interaction between the Tb-5d electrons, with $J=0$ for sake of simplicity. We found that the energy difference between the \texttt{CUBIC} and \texttt{HUND} states remains basically unchanged.

\subsection{Spin-orbit coupling and magnetic properties}

In this subsection we will analyze the influence of spin-orbit coupling and the magnetic properties. Before we continue two things must be emphasized. First, we used the equilibirum structures obtained above (see subsection Lattice properties) for this investigation.  Second, all LDA/GGA as well as LDA+U and LDA+DMFT[HIA] calculations reported above are done without spin-orbit coupling. However, for the magnetic properties to which the 4f-electron contribution is crucial, the spin-orbit coupling must be included. Note that the orbital moments discussed in the previous subsection for the LDA+U approach were purely induced by the local Coulomb interaction, which can favour states obeying the second Hund's rule~\cite{solovyev98prl80:5758}. The inclusion of the spin-orbit coupling, instead, offers a more complete picture and allows us to also consider the effects associated to the third Hund's rule. The results of our calculations, for selected methods, are summarized in Table~\ref{magneticmoments}. For DFT simulations in LDA and 4f-electrons treated as valence states, a total moment of 7.3 $\mu_B$, consisting of a spin moment of 6 $\mu_B$ and an orbital moment of 1.3 $\mu_B$, is found. The self-consistent LDA+U simulations were started from the \texttt{CUBIC} and \texttt{HUND} states discussed previously, and are therefore indicated with the same labels, although the cubic symmetry is now broken due to presence of spin-orbit coupling and finite magnetization. 
\begingroup
\squeezetable
\begin{table}[bt]
\begin{tabular}{|l|c|c|c|}
\hline
& & & \\[-2.2ex]
\textbf{Method} & $L_z$ & $2S_z$ & $L_z+2S_z$ \\
\hline
& & & \\[-2.2ex]
LDA \texttt{VALENCE} & 1.3 & 6.0 & 7.3 \\
\hline
& & & \\[-2.2ex]
LDA+U \texttt{CUBIC} & 2.2 & 5.3 & 7.5 \\
\hline
& & & \\[-2.2ex]
LDA+U \texttt{HUND} & 3.4 & 5.0 & 8.4 \\
\hline
& & & \\[-2.2ex]
LDA+DMFT[HIA] & 2.7 & 5.7 & 8.4 \\
\hline
& & & \\[-2.2ex]
Experiment & -- & -- & 8.5 \\
\hline
\end{tabular}
\caption{Calculated and experimental values for the orbital, spin and total magnetic moments of TbN-bulk. The meaning of the labels is the same as in Table~\ref{structure1}, but here we have also included corrections due to the spin-orbit coupling. The experimental value is taken from the study of Ref.~\onlinecite{wachter98ssc105:675}, as discussed in the main text.}
\label{magneticmoments}
\end{table}
\endgroup
When starting from the \texttt{CUBIC} state, we obtain a spin moment of 5.3 $\mu_B$ and an orbital moment of 2.2 $\mu_B$, giving a total moment of 7.5 $\mu_B$. Conversely, when starting from the \texttt{HUND} state, we obtain a spin moment of 5.0 $\mu_B$ and an orbital moment of 3.4 $\mu_B$. These new simulations can also be used to check the previously discussed total energies of the LDA+U ground state. With the inclusion of relativistic effects, we find that the \texttt{CUBIC} state is 0.29 eV lower in energy than the \texttt{HUND} state, in qualitative agreement with our previous results. These results are also presented in Figure~\ref{Scheme}.
 
Evaluating the magnetic moment with LDA+DMFT[HIA] is a bit more involved, due to problems related to the double counting correction~\cite{granas12cms55:295}. The HIA requires as input the projected local Hamiltonian of the 4f shell, which, for spin-polarized solutions, contains the 4f-shell exchange splitting. This exchange splitting arises from both intra-orbital and inter-orbital contributions~\cite{granas12cms55:295}. The former is due to the local Coulomb interaction between the 4f electrons, and should ideally be considered only at the level of the HIA. Therefore, one should remove it from the input local Hamiltonian, but unfortunately it is not possible to disentangle this term from the inter-orbital contributions. Here we solve this problem by substituting the entire exchange splitting with an approximate expression for the inter-orbital contributions, as is explained below.

The exchange energy of rare-earths can be approximated~\cite{white:magnetism, brooks89jpcm1:5861} as 
\begin{equation}
E_{X}=\frac{1}{4}\sum_{l,l'}I_{ll'}m_{l}m_{l'}.
\label{EX}
\end{equation}
Here $l$ denotes the angular quantum number, $m_{l}=n_{l}^{\uparrow}-n_{l}^{\downarrow}$ are the corresponding spin-moments and $I_{ll'}$ are atomic exchange integrals. Since the s and p states do not create any significant magnetic moment, the main inter-orbital contribution to the exchange energy of the 4f-states comes from the interaction with the Tb 5d-states. Thus, the exchange splitting of the 4f shell caused by the interaction with the d-states can be calculated from the inter-orbital energy \textit{$E_{X}^{fd}={I_{fd}m_{f}m_{d} / 2}$} as follows:
\begin{equation}
\begin{split}
\Delta E_{X}^{fd} & =\frac{\partial E_{X}^{fd}}{\partial n_{f}^{\uparrow}}-\frac{\partial E_{X}^{fd}}{\partial n_{f}^{\downarrow}} \\
 & =\frac{\partial E_{X}^{fd}}{\partial m_{f}} \frac{\partial m_{f}}{\partial n_{f}^{\uparrow}}-\frac{\partial E_{X}^{fd}}{\partial n_{f}^{\downarrow}} \frac{\partial m_{f}}{\partial n_{f}^{\downarrow}}=I_{fd}m_{d}.
\end{split}
\label{EXsplitting}
\end{equation}
This exchange-interaction acts as an effective field on the 4f shell, and we have added it as such, with a strength determined by Eq.~(\ref{EXsplitting}). This was evaluated from an $I_{fd}$ integral of 7 mRy, taken from Ref.~\onlinecite{brooks89jpcm1:5861}, and a self-consistently calculated value of $m_{d}$ of 0.006 $\mu_B$. This exchange interaction is then considered as an effective field, which breaks the 2J+1 degeneracy of the ground state configuration, so that only the lowest $|{J,M_J}\rangle$ level is occupied. We find that this level (which does hybridize slightly with other orbitals) carries a magnetic moment of 8.4 $\mu_B$, of which 2.7 $\mu_B$ comes from the orbital part and 5.7 $\mu_B$ from the spin part. The calculation of the \textit{$m_{d}$} moment is associated with some uncertainty, since this value will depend slightly on details of the calculation, e.g. the choice of muffin-tin radius. To test the sensitivity of the calculated 4f moment to the value of \textit{$m_{d}$}, we increased \textit{$m_{d}$} by one order of magnitude in Eq.~(\ref{EXsplitting}), and performed a calculation of the 4f moment as described above. We then obtain a 4f projected moment of 8.7 $\mu_B$, of which 2.8 $\mu_B$ comes from the orbital part and 5.9 $\mu_B$ from the spin part. Hence the sensitivity of the 4f moment to the choice of the parameters in Eq.~(\ref{EXsplitting}) is not large, and the important aspect is that the 2J+1 degeneracy is lifted by the interaction with an inter-orbital exchange field.

From the low temperature experimental work in Ref.~\onlinecite{hulliger_hpcre_79,wachter98ssc105:675} it has been reported that the magnetic ordering of TbN, and other rare-earth nitrides, depends critically on the carrier concentration, which can be controlled by slight modifications of N concentration. Saturation moments of 6.7-7 $\mu_B$/Tb atom have been reported for samples where there is still a small antiferromagnetic component~\cite{wachter98ssc105:675} to the essentially dominating ferromagnetic exchange. Samples that have solely ferromagnetic inter-atomic exchange have been reported to have moments of 8.5 $\mu_B$/Tb atom.~\cite{wachter98ssc105:675} This value is close to the value expected from the Standard Model of a trivalent Tb atom, and is also close to the calculations based on LDA+U \texttt{HUND} and LDA+DMFT[HIA]. The latter, however, agrees better with the Standard Model with respect to the balance between spin and orbital contributions to the total magnetic moment. Namely from the standard model an orbital momentum contribution of 3 $\mu_B$/atom and a spin moment of 6 $\mu_B$/atom is expected.

\subsection{Spectral properties}
\begin{figure*}[t]
\includegraphics[trim=0.1 360.1 0.1 0.1,clip,width=16cm]{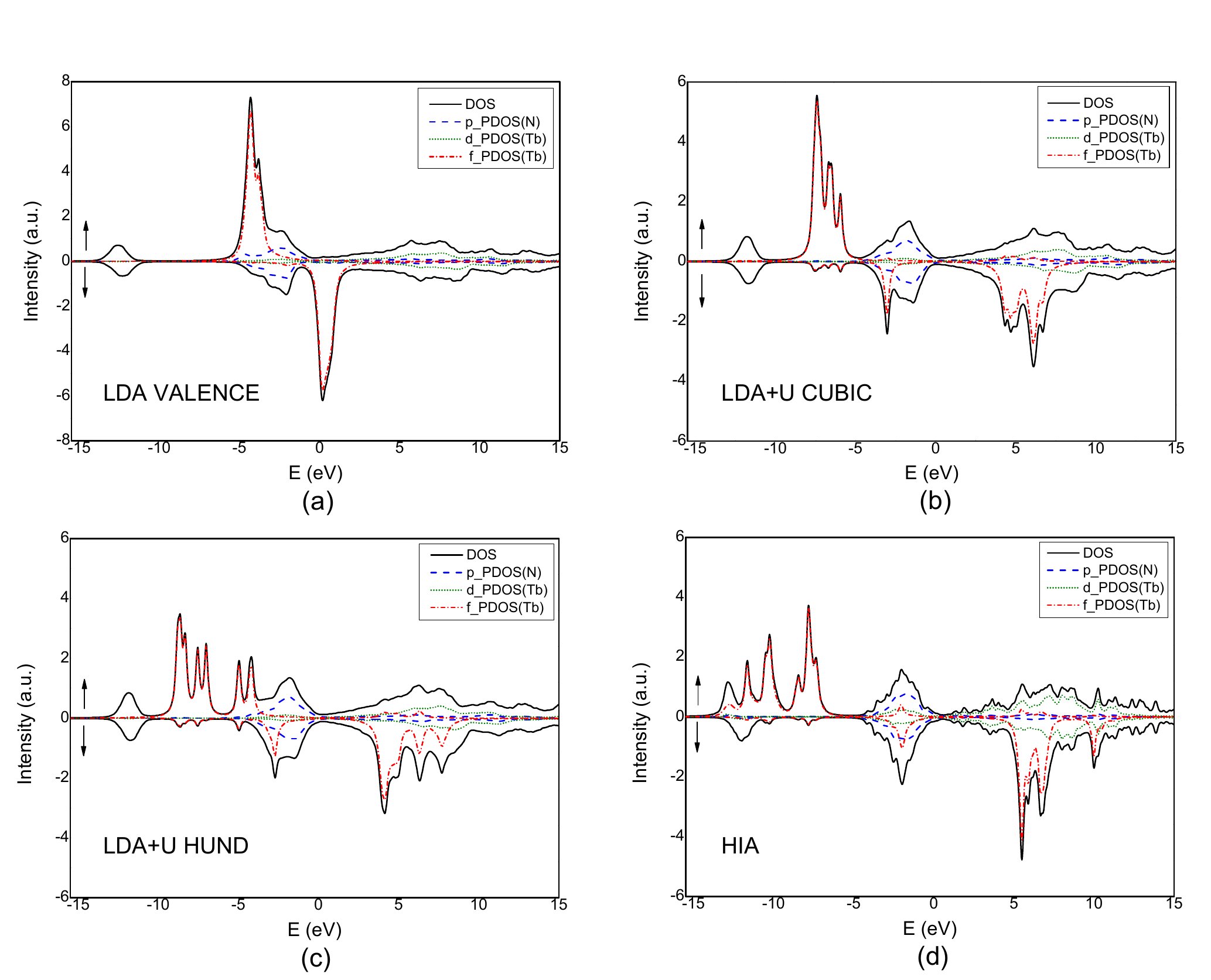}
\caption{Total density of states (full black lines), and projected density of states of N-2p (dashed blue lines), Tb-5d (dotted green lines) and Tb-4f (dashed-dotted red lines) electrons. The most relevant methods of this study are reported in the different quadrants: LDA \texttt{VALENCE} (a), LDA+U \texttt{CUBIC} (b), LDA+U \texttt{HUND} (c) and LDA+DMFT[HIA] (d). Note that the scale of the y-axis is different in (a) compared to (b), (c) and (d).} 
\label{Spectral}
\end{figure*}

In Fig.~\ref{Spectral} we show the total density of states and the projected density of states for the N-2p, Tb-5d and Tb-4f electrons. We report on all the methods discussed in the previous subsection, i.e. LDA \texttt{VALENCE} (Fig.~\ref{Spectral}a), LDA+U \texttt{CUBIC} (Fig.~\ref{Spectral}b), LDA+U \texttt{HUND} (Fig.~\ref{Spectral}c) and LDA+DMFT[HIA] (Fig.~\ref{Spectral}d). All these calculations are spin-polarized and include the effects due to the spin-orbit coupling. For LDA \texttt{VALENCE}, two sharp peaks are observed in the 4f-projected density of states, one corresponding to the majority spin channel and the other to the minority spin channel. This minority spin channel is pinned at the Fermi level, because it is partially filled. At the moment no experimental photoemission spectra of TbN are available to compare with theoretical spectra. However, due to the highly localized character of the 4f-electrons, it is very unlikely that density of states can have a finite Tb-4f contribution at Fermi level. In trivalent elemental Tb, where several material properties emphasize a smaller degree of localization, the 4f spectral features are found at higher binding energy~\cite{svane06ssc140:364,lebegue05prb72:245102}.

In Fig.~\ref{Spectral}b we see that for the LDA+U \texttt{CUBIC} solution, which is our LDA+U ground-state, there is no or little 4f spectral intensity at the Fermi level. We observe instead different peaks of the 4f-projected density of states well below and well above the Fermi level. Here the peaks at -8 eV and -7 eV come from respectively the \textit{$t_{1u}$} and \textit{$t_{2u}$} state, and the peak at -6 eV from the \textit{$a_{2u}$} state.  These peaks are not due to the formation of atomic multiplets, but are caused on a single particle level. Hence, although they have more structure, compared to the LDA calculation, these structures are not the ones typically found for trivalent Tb, in elemental form or in compounds.

\begin{figure*}[bt]
\includegraphics[width=8cm]{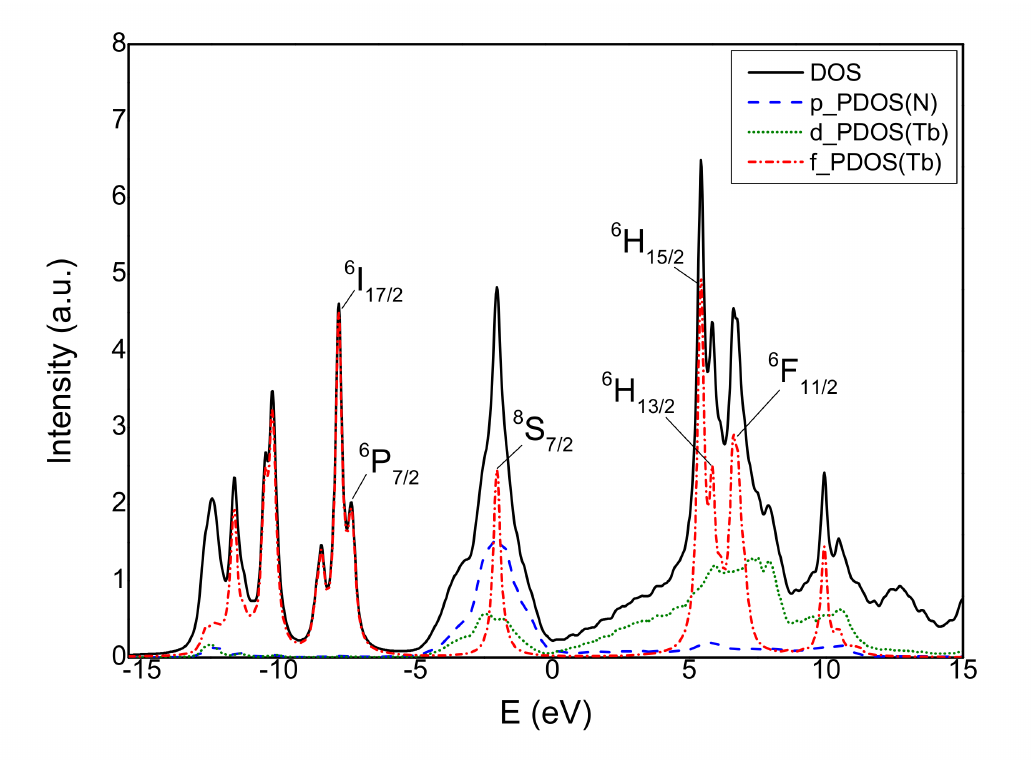}
\caption{Total density of states (full black lines), and projected density of states of N-2p (dashed blue lines), Tb-5d (dotted green lines) and Tb-4f (dashed-dotted red lines) electrons in LDA+DMFT[HIA] without spin-polarization.} 
\label{HIA-unpol}
\end{figure*}

In Fig.~\ref{Spectral}c we report the spectrum of the LDA+U \texttt{HUND} solution. This is not our ground-state but it may be useful to look at its spectral properties in order to check if the observed features resemble or not the atomic-like multiplets. To this aim we can compare Fig.~\ref{Spectral}c and Fig.~\ref{Spectral}d. In the LDA+DMFT[HIA] calculations, the 4f-projected density of states (dashed-dotted red line) undoubtedly shows peaks caused by the formation of atomic multiplets~\cite{pourovskii09prl102:096401,thunstrom09prb79:165104}. The spectral structure below the Fermi energy corresponds to \textit{$f^8$} to \textit{$f^7$} transitions, while the structure above it to \textit{$f^8$} to \textit{$f^9$} transitions. 
The largest differences between LDA+U \texttt{HUND} and LDA+DMFT[HIA] calculation can be found in the majority spin channel. In the LDA+U \texttt{HUND} spectrum the 4f-peaks are closer to the Fermi level, of about 4 eV, and also the shape and relative positions of the peaks seem to differ. For example, LDA+U HUND has two 4f-peaks at -5 and -4 eV, which are absent in the LDA+DMFT[HIA] spectrum. Also LDA+DMFT[HIA] has 4f-peaks with multiplet features below -10 eV, while LDA+U HUND does not have this.

Due to that some majority 4f-states overlap with the N-2p states, a (small) hybridization with them can also influence the binding properties (see again Table~\ref{structure1}). One could speculate that these differences are caused by an artificial increase of the exchange splitting due to the method illustrated in the previous subsection. To verify this point, we have computed the spectral properties also in the paramagnetic phase, shown in Fig.~\ref{HIA-unpol}. The Hubbard I approximation is a proper many-body theory, and takes into account several Slater determinants in the ground state and excited states. Therefore the paramagnetic spectrum is expected to be very similar to the spin-integrated ferromagnetic spectrum. However, in the paramagnetic phase no approximation on the exchange has been made, and therefore eventual differences with the magnetic case could be traced to that. The total densities of states DOS in Fig.~\ref{Spectral}d and Fig.~\ref{HIA-unpol} are very similar, confirming that, in TbN, the differences between LDA+U and LDA+DMFT[HIA] are indeed fundamental. The qualitative differences outlined in this paragraph are in good agreement with a previous study on ErAs, where similar methods were employed~\cite{pourovskii09prl102:096401}. However, in the latter study the largest discrepancies between LDA+DMFT[HIA] and LDA+U \texttt{HUND} were found in the minority spin channel.
 
Finally in Fig.~\ref{HIA-unpol} the major excitation peaks were also labelled in the corresponding atomic notation. The first peak below the Fermi level, at around -4 eV, corresponds to a transition to the \textit{$^8S_{7/2}$} state. The first peak above the Fermi level, at around 3 eV, corresponds to a transition to \textit{$^6H_{15/2}$}. Overall, the spectra of Fig.~\ref{Spectral}d and Fig.~\ref{HIA-unpol} are consistent with a typical spectrum of a trivalent Tb atom, either in elemental form or in compounds~\cite{svane06ssc140:364,lebegue05prb72:245102,lebegue06jpcm18:6329}. Besides the obvious advantage that multiplet-configurations are taken into account in the LDA+DMFT[HIA] scheme, we also expect from previous calculations on heavy rare-earth elements that the LDA+DMFT[HIA] calculation will resemble the measured spectral properties of TbN-bulk best (see e.g. \onlinecite{svane06ssc140:364, lebegue05prb72:245102,lebegue06jpcm18:6329}). In these works an excellent comparison is found between LDA+DMFT[HIA] calculated and experimental (XPS and BIS) spectra, also including elemental, trivalent Tb.

\section{Conclusion}
We have investigated the applicability of several theoretical methods to describe the 4f states of an archetypical rare-earth compound, TbN. These treatments included LDA/GGA (with 4f-electrons in valence and core), LDA/GGA+U and LDA+DMFT in the Hubbard-I approximation. We have focused our investigation on structural properties, equilibrium lattice constant, bulk modulus, magnetism and spectra. We have studied two significant local minima of the LDA+U method. One is characterized by a 4f density matrix close to that given by Hund's rules, and labelled as \texttt{HUND}. The other one, labelled as \texttt{CUBIC}, originates from the one-particle levels of a cubic crystal field, and retains the cubic symmetry when spin-orbit coupling is neglected. This \texttt{CUBIC} solution has been found to have lower energy compared to the \texttt{HUND} solution  in all cases, i.e. with and without spin-orbit coupling, with and without considering a $U_\text{d}$ term for the Tb-5d states, and also with a different electronic structure code. 

When focusing on the equilibrium lattice constant, all methods reproduce the measured data with good accuracy, except for LDA with 4f-electrons in the valence. The bulk modulus and shear constant appear to be rather sensitive to the method used, and we find that the LDA+U method with a \texttt{HUND} solution results in a negative $C^{\prime}$ constant, which is the signature of a sizeable tetragonal distortion of the NaCl-structure. This result is, however, in contradiction to experiments. In case of the magnetic properties only LDA+DMFT in the Hubbard I approximation is consistent with the Standard Model of the rare-earths, and gives a total, as well as spin and orbital, magnetic moment in good agreement with experiment, while all other methods have major or minor deficiencies. For the spectral properties only LDA+DMFT in the Hubbard I approximation was able to capture the expected atomic multiplets, but our assessment cannot be complete due to the lack of experimental photoemission data. 

Thus, our overall conclusion is that of all the theoretical methods used for the calculation of the different physical properties of TbN, it is only LDA+DMFT that is consistent with the Standard Model and available experimental data. This conclusion is expected to hold for rare-earth systems in general, and it is suggested here that for theoretical studies of rare-earth systems, the LDA+DMFT in the Hubbard-I approximation should be considered as the primary theoretical tool.

\begin{acknowledgments}
We acknowledge support from the Swedish Research Council (VR),  eSSENCE, STANDUPP, and the Swedish National Allocations Committee (SNIC/SNAC). O.E. also acknowledges support from ERC (project 247062 - ASD) and the KAW foundation. Also the Nederlandse Organisatie voor Wetenschappelijk Onderzoek (NWO) is acknowledged for giving us a grant on the LISA supercomputer. Also SURFsara (www.surfsara.nl) is acknowledged for the usage of LISA and their support. M. I. K. acknowledges support from ERC (project 338957 - FEMTO/NANO).
\end{acknowledgments}

\bibliography{strings,kylie}

\end{document}